\documentclass[sn-nature]{sn-jnl}

\newcommand\aj{AJ}\newcommand\araa{ARA\&A}\newcommand\apj{ApJ}\newcommand\apjl{ApJL}     \newcommand\apjs{ApJS}\newcommand\aap{A\&A}  \newcommand\mnras{MNRAS}\newcommand\nat{Nature}

\newcommand\pasa{PASA}

\usepackage{graphicx}\usepackage{multirow}\usepackage{amsmath,amssymb,amsfonts}\usepackage{amsthm}\usepackage{mathrsfs}\usepackage[title]{appendix}\usepackage{xcolor}\usepackage{textcomp}\usepackage{manyfoot}\usepackage{booktabs}\usepackage{algorithm}\usepackage{algorithmicx}\usepackage{algpseudocode}\usepackage{listings}\usepackage{lineno}
\usepackage{natbib}
\usepackage{xspace}

\newcommand{\HI}{H\,{\textsc{i}}\xspace}
\newcommand{\lcdm}{{\ensuremath{\Lambda}CDM}\xspace}

\newcommand{\sref}{\nameref} 

\renewcommand{\orcidlogo}{\includegraphics[width=10pt]{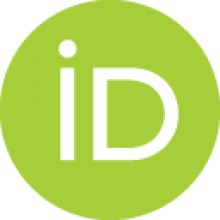}}
\renewcommand{\orcid}[1]{\href{https://orcid.org/#1}{\orcidlogo}}

\renewcommand{\citet}[1]{Ref.\citep{#1}}

\title[Massive Galaxy Halos Contain Less Inner Dark Matter Than Predicted]{Massive Galaxy Halos Contain Less Inner Dark Matter Than Predicted\footnotemark[2]
}

\author[1, 2]{\fnm{Yu-Chen}  \sur{Wang}\,\orcid{0000-0002-8429-7088}}
\author*[1, 2]{\fnm{Yingjie}  \sur{Peng}\,\orcid{0000-0003-0939-9671}}\email{yjpeng@pku.edu.cn} 
\author[3, 4]{\fnm{Xiaohu}  \sur{Yang}\,\orcid{0000-0003-3997-4606}}
\author[2, 1]{\fnm{Luis C.}  \sur{Ho}\,\orcid{0000-0001-6947-5846}}
\author[1, 2]{\fnm{Dingyi}  \sur{Zhao}\,\orcid{0009-0001-1564-3944}}
\author[5]{\fnm{Jing}  \sur{Dou}\,\orcid{0000-0002-6961-6378}}
\author[6, 7]{\fnm{Hao}  \sur{Fu}\,\orcid{0009-0002-8051-1056}}
\author[1, 2]{\fnm{Zeyu}  \sur{Gao}\,\orcid{0000-0002-0182-1973}}
\author[8]{\fnm{Qiusheng}  \sur{Gu}\,\orcid{0000-0002-3890-3729}}
\author[2]{\fnm{Fangzhou}  \sur{Jiang}\,\orcid{0000-0001-6115-0633}}
\author[1, 2]{\fnm{Yukun}  \sur{Liu}}
\author[9, 10, 11]{\fnm{Roberto}  \sur{Maiolino}\,\orcid{0000-0002-4985-3819}}
\author[12]{\fnm{Houjun}  \sur{Mo}\,\orcid{0000-0001-5356-2419}}
\author[1, 2]{\fnm{Canpo}  \sur{Su}}
\author[13]{\fnm{Bitao}  \sur{Wang}\,\orcid{0000-0002-6137-6007}}
\author[14, 15]{\fnm{Kai}  \sur{Wang}\,\orcid{0000-0002-3775-0484}}
\author[2]{\fnm{Bingxiao}  \sur{Xu}}
\author[6]{\fnm{Feng}  \sur{Yuan}\,\orcid{0000-0003-3564-6437}}
\author[1, 2]{\fnm{Kunyao}  \sur{Zhao}\,\orcid{0009-0000-3647-6527}}
\author[1, 2]{\fnm{Xingye}  \sur{Zhu}\,\orcid{0000-0002-9529-1044}}

\affil[1]{\orgdiv{Department of Astronomy, School of Physics}, \orgname{Peking University}, \orgaddress{\street{5 Yiheyuan Road},  \city{Beijing} \postcode{100871}, \country{People's Republic of China}}}
\affil[2]{\orgdiv{Kavli Institute for Astronomy and Astrophysics}, \orgname{Peking University}, \orgaddress{\street{5 Yiheyuan Road},  \city{Beijing} \postcode{100871}, \country{People's Republic of China}}}
\affil[3]{\orgdiv{Tsung-Dao Lee Institute, and Shanghai Key Laboratory for Particle Physics and Cosmology}, \orgname{Shanghai Jiao Tong University}, \orgaddress{\city{Shanghai} \postcode{200240}, \country{People's Republic of China}}}
\affil[4]{\orgdiv{Department of Astronomy, School of Physics and Astronomy}, \orgname{Shanghai Jiao Tong University}, \orgaddress{\city{Shanghai} \postcode{200240}, \country{People's Republic of China}}}
\affil[5]{\orgdiv{National Astronomical Observatories, Chinese Academy of Sciences},  \orgaddress{\city{Beijing} \postcode{100101}, \country{People's Republic of China}}}
\affil[6]{\orgdiv{Center for Astronomy and Astrophysics and Department of Physics}, \orgname{Fudan University}, \orgaddress{\city{Shanghai} \postcode{200438}, \country{People's Republic of China}}}
\affil[7]{\orgdiv{School of Physics and Astronomy}, \orgname{University of Southampton}, \orgaddress{\street{Highfield},  \city{Southampton} \postcode{SO17 1BJ}, \country{UK}}}
\affil[8]{\orgdiv{School of Astronomy and Space Science}, \orgname{Nanjing University}, \orgaddress{\city{Nanjing} \postcode{210093}, \country{People's Republic of China}}}
\affil[9]{\orgdiv{Cavendish Laboratory}, \orgname{University of Cambridge}, \orgaddress{\street{19 J.J. Thomson Avenue},  \city{Cambridge} \postcode{CB3 0HE}, \country{UK}}}
\affil[10]{\orgdiv{Kavli Institute for Cosmology}, \orgname{University of Cambridge}, \orgaddress{\street{Madingley Road},  \city{Cambridge} \postcode{CB3 0HA}, \country{UK}}}
\affil[11]{\orgdiv{Department of Physics and Astronomy}, \orgname{University College London}, \orgaddress{\street{Gower Street},  \city{London} \postcode{WC1E 6BT}, \country{UK}}}
\affil[12]{\orgdiv{Department of Astronomy}, \orgname{University of Massachusetts}, \orgaddress{\city{Amherst}, \state{MA} \postcode{01003}, \country{USA}}}
\affil[13]{\orgdiv{School of Physics and Electronics}, \orgname{Hunan University}, \orgaddress{\city{Changsha} \postcode{410082}, \country{People's Republic of China}}}
\affil[14]{\orgdiv{Institute for Computational Cosmology, Department of Physics}, \orgname{Durham University}, \orgaddress{\street{South Road},  \city{Durham} \postcode{DH1 3LE}, \country{UK}}}
\affil[15]{\orgdiv{Centre for Extragalactic Astronomy, Department of Physics}, \orgname{Durham University}, \orgaddress{\street{South Road},  \city{Durham} \postcode{DH1 3LE}, \country{UK}}}

\abstract{
The mass profiles of galaxy halos encode how baryons reshape dark matter distribution, yet direct observational constraints across the full radial range remain scarce. Here we combine stellar kinematics from MaNGA, \HI dynamical measurements from ALFALFA, and independently calibrated halo masses of SDSS groups to statistically reconstruct the mass distribution of central galaxies over nearly two orders of magnitude in radius. We demonstrate that \HI data alone do not provide reliable total halo mass estimates, necessitating an independent group-based halo-mass scale. Compared to the IllustrisTNG and EAGLE simulations, the observational profiles of low-mass halos are broadly consistent; in contrast, massive observed halos exhibit systematically lower dynamical masses at the \HI radius, lower inner dark-matter masses, and lower central dark-matter fractions 
(about 4$\sigma$ difference in units of population scatter) at fixed total halo mass. After subtracting baryonic contributions, the inferred dark-matter profiles remain broadly consistent with an NFW form, but with lower effective concentrations than predicted for massive halos. These results suggest that the inner dark-matter content of massive halos has been reduced more significantly than predicted by current hydrodynamical simulations, plausibly due to long-term baryonic halo heating in massive systems.
}

\begin{document}

\maketitle

\renewcommand{\thefootnote}{\fnsymbol{footnote}}
\footnotetext[2]{This is the 13th paper in the ``From Halos to Galaxies'' series.}
\renewcommand{\thefootnote}{\arabic{footnote}}

In the standard \lcdm paradigm of galaxy formation and evolution, galaxies form from gas accreted into the centers of dark matter halos \citep{Mo:2010gfe..book.....M}. 
Baryonic processes associated with galaxies can, in turn, alter the distribution of dark matter \citep[e.g.][]{Schaller:2015MNRAS.452..343S,Schaller:2015MNRAS.451.1247S,Anbajagane:2022MNRAS.509.3441A,Sorini:2025MNRAS.536..728S,Gebhardt:2026arXiv260106258G}, causing deviations from
predictions made by dark-matter-only $N$-body simulations.
The baryonic effects 
are considered solutions to
the \lcdm's small-scale challenges \citep{Weinberg:2015PNAS..11212249W, Bullock:2017ARA&A..55..343B},
such as 
the cusp--core problem \citep[e.g.][]{Moore:1994Natur.370..629M, Oh:2011AJ....141..193O, Governato:2012MNRAS.422.1231G, Martizzi:2013MNRAS.432.1947M}. 
Common baryonic processes include
adiabatic contraction \citep{Blumenthal:1986ApJ...301...27B, Gnedin:2004ApJ...616...16G}, where dark matter halos become more concentrated due to the collapse and accumulation of baryons in the centers of halos, and baryonic halo heating, caused by feedbacks from stars and active galactic nuclei (AGN) and merger- or satellite-induced dynamical friction heating
\citep[e.g.][]{Pontzen:2012MNRAS.421.3464P,El-Zant:2016MNRAS.461.1745E,Freundlich:2020MNRAS.491.4523F,Dekel:2021MNRAS.508..999D,Li:2023MNRAS.518.5356L,Arora:2024MNRAS.529.2047A}, 
reducing the central dark matter content.
Consequently, constraints on the profiles (i.e., the mass distribution) of dark-matter halos
provide a direct test of how baryonic processes reshape dark matter.

The observational constraints on halo profiles, however, remain fragmented, because different tracers probe different radial regimes and are rarely tied to a common halo-mass scale. Stellar kinematic observations constrain the baryon-dominated inner galaxy, typically covering the central 1--2 effective radii ($R_\mathrm{e}$). \HI observations probe larger radii, but are still far inside the virial boundary of halos, necessitating the incorporation of independent information at the halo scale. To connect these scales, we combine the stellar kinematics from the Mapping Nearby Galaxies at
Apache Point Observatory (MaNGA) survey \citep{Bundy:2015ApJ...798....7B, Abdurro'uf:2022ApJS..259...35A}, \HI measurements from the Arecibo Legacy Fast ALFA (ALFALFA) survey \citep{Haynes:2011AJ....142..170H, Haynes:2018ApJ...861...49H}, and independently calibrated SDSS group halo masses \citep{Zhao:2025ApJ...979...42Z}, enabling a statistical reconstruction of galaxy mass profiles from the inner galaxy to the halo scale, spanning nearly two orders of magnitude in radius. The three datasets provide complementary radial anchors.

The data we use can be expressed as dynamical masses probed at certain radii.
For MaNGA data, we adopt the enclosed dynamical mass, $M_\mathrm{dyn}(<r_{1/2})$, and dark matter mass, $M_\mathrm{DM}(<r_{1/2})$, at the 3D half-light radius $r_{1/2}$, inferred by Jeans Anisotropic Modelling (JAM) \citep{Cappellari:2008MNRAS.390...71C, Cappellari:2020MNRAS.494.4819C} in the MaNGA DynPop project \citep{Zhu:2023MNRAS.522.6326Z}. 
The spatially integrated \HI 21-cm line data from ALFALFA provide the dynamical mass $M_\mathrm{dyn}(<r_\mathrm{HI})$ at the \HI radius $r_\mathrm{HI}$, derived in \citet{Yu:2022ApJS..261...21Y} from 
the \HI mass $M_\mathrm{HI}$, line widths, 
and the tight \HI mass--size relation \citep{Wang:2016MNRAS.460.2143W}.
Since $r_\mathrm{HI}$ remains several times smaller than the halo virial radius $r_\mathrm{h}$, it is essential to incorporate independent estimates of total halo masses $M_\mathrm{h}$, i.e., the mass enclosed within $r_\mathrm{h}$.
We adopt the improved $M_\mathrm{h}$ estimates in \citet{Zhao:2025ApJ...979...42Z},
obtained with a machine learning model and validated against both the halo mass function and weak lensing stellar-to-halo mass relations.
We restrict our study to central galaxies, as the $M_\mathrm{h}$ estimates do not include subhalo masses of satellite galaxies.
For brevity, we refer to the aforementioned three datasets 
as the MaNGA, \HI, and $M_\mathrm{h}$ data.
Details on data processing and selections are described in \sref{sec:method}.

\begin{figure*}
\centering
\includegraphics[width=.8\linewidth]{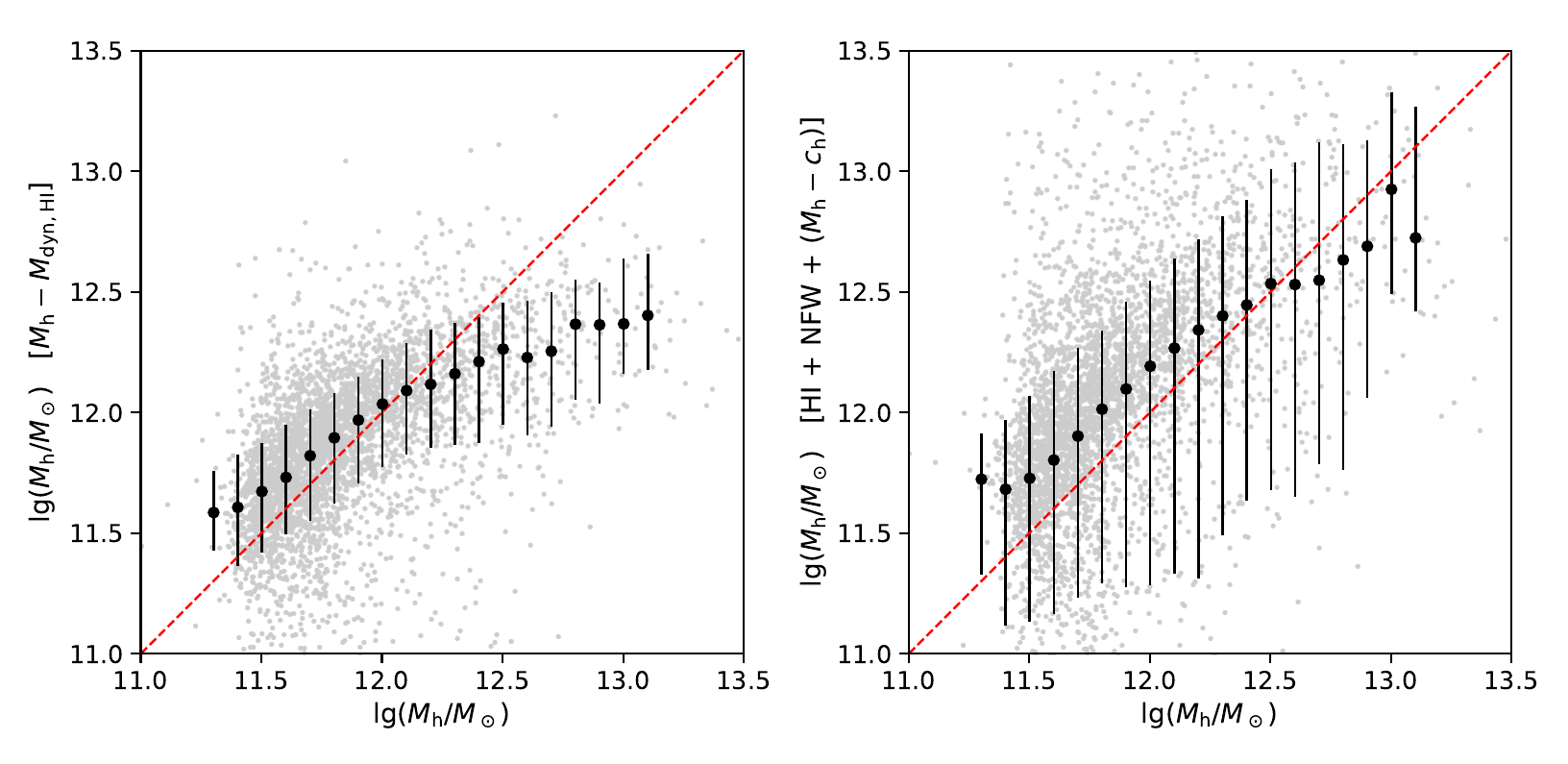}
\caption{\textbf{Comparison of SDSS group halo masses ($M_\mathrm{h}$) 
and  those solely from} \HI \textbf{data.} The x-axes of both panels show the halo mass $M_\mathrm{h}$ estimated from group properties using machine learning \citep{Zhao:2025ApJ...979...42Z}, and y-axes show estimations of $M_\mathrm{h}$ from \HI using different methods. The left panel shows $M_\mathrm{h}$ directly converted from the dynamical mass, $M_\mathrm{dyn}(<r_\mathrm{HI})$, using an empirical correlation \citep{Yu:2020ApJ...898..102Y}; 
the right panel shows $M_\mathrm{h}$ calculated from $(r_\mathrm{HI}, M_\mathrm{dyn}(<r_\mathrm{HI}))$ 
by assuming
an NFW profile and a correlation between its two parameters, $M_\mathrm{h}$ and concentration $c_\mathrm{h}$ (see \sref{sec:method} for details). Only central galaxies are included, and individual galaxies are marked with the gray dots. The medians and 1-sigma ranges of sequential $M_\mathrm{h}$ bins are marked with the black dots and errorbars. The red dashed lines represent the line where the x- and y-values are equal.
It can be seen that estimating the total halo mass $M_\mathrm{h}$ directly from the \HI radii and dynamical masses results in a significant scatter or systematic error.} \label{fig:comphihalomass}
\end{figure*}

The total halo mass $M_\mathrm{h}$ cannot be robustly inferred from \HI alone, making our external halo mass anchor essential. 
Fig.\ \ref{fig:comphihalomass}
compares the 
SDSS group
halo mass $M_\mathrm{h}$ 
\citep{Zhao:2025ApJ...979...42Z} (our fiducial halo mass)
with alternative halo mass estimations derived solely from the \HI data. For these alternative estimations, the left panel
directly converts
the \HI dynamical mass, $M_\mathrm{dyn}(<r_\mathrm{HI})$, to $M_\mathrm{h}$ using an empirical correlation \citep{Yu:2020ApJ...898..102Y} calibrated from resolved \HI observations. 
The right panel uses both $M_\mathrm{dyn}(<r_\mathrm{HI})$ and the radius $r_\mathrm{HI}$ to derive the total halo mass, assuming an Navarro–Frenk–White (NFW) profile \citep{Navarro:1996ApJ...462..563N} and a correlation between the two NFW parameters, i.e., $M_\mathrm{h}$ and concentration $c_\mathrm{h}$. 
As seen, \HI-based halo masses are broadly acceptable at low halo mass, but become systematically underestimated at the massive end, or highly scattered when an NFW prior is imposed.
Details on the above approaches to derive $M_\mathrm{h}$ are given in \sref{sec:method}.

\begin{figure*}
\centering
\includegraphics[width=.8\linewidth]{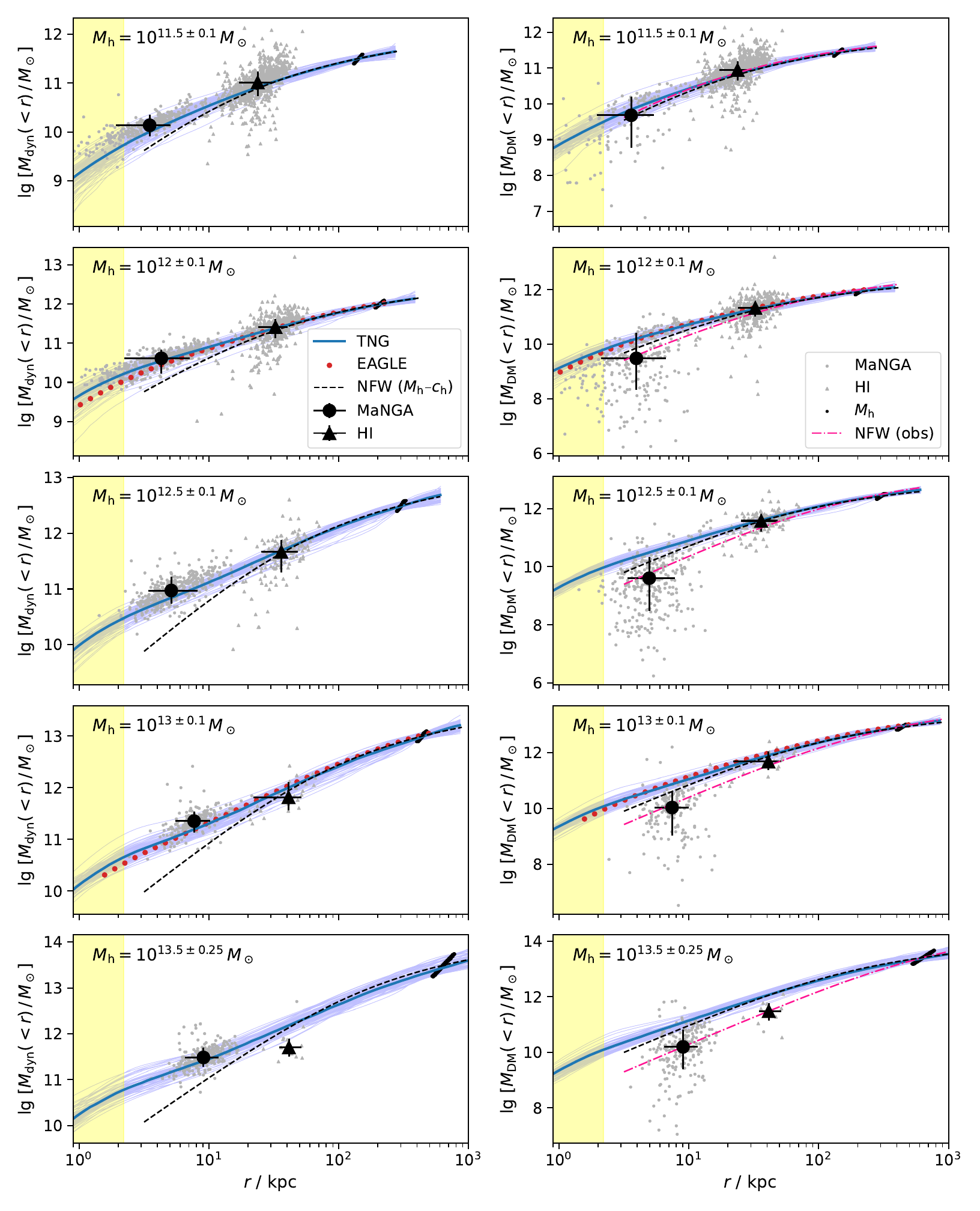}
\caption{\textbf{The dynamical mass $M_\mathrm{dyn}(<r)$ and dark matter mass $M_\mathrm{DM}(<r)$ profiles.} The total dynamical masses (left panels) and dark matter masses (right panels) enclosed within spheres of radii $r$ are plotted as functions of $r$. Each row of panels shows a different halo mass $M_\mathrm{h}$ bin, with the ranges indicated at the top left corners. The blue thick lines indicate the median mass profiles of the TNG-100 simulation, and the corresponding thin lines plot random samples of individual halos, hinting the scatter of the population. The yellow backgrounds mark the regions of $r\lesssim 2.2\, \mathrm{kpc}$, approximately 3 times the TNG-100 softening length, where the TNG results should be considered with caution. The red dots in $M_\mathrm{h}\sim 10^{12}M_\odot$ or $10^{13}M_\odot$ show the profiles of stacked halos in the EAGLE simulation, reproduced from Fig.~6 in \citet{Schaller:2015MNRAS.451.1247S}. The observational data of enclosed dynamical or dark matter masses are marked with gray triangles (\HI) and dots (MaNGA). The large black dots and triangles mark the corresponding median values of the data, with the errorbars showing the 1-sigma quantiles. 
The little black dots 
mark the virial radii $r_\mathrm{h}$ and the enclosed masses 
from the $M_\mathrm{h}$ data
($M_\mathrm{dyn}(<r_\mathrm{h})=M_\mathrm{h}$, and $M_\mathrm{DM}(<r_\mathrm{h})$ estimated based on the cosmic baryon fraction),
and thus indicating the sizes of the halo mass bins. 
The black dashed lines show the NFW mass profiles for the corresponding $M_\mathrm{h}$ and the concentration parameters $c_\mathrm{h}$ calculated using the theoretical $M_\mathrm{h}$--$c_\mathrm{h}$ relation from $N$-body (i.e., dark-matter-only) simulations \citep{Ishiyama:2021MNRAS.506.4210I}. 
The pink dash-dotted lines indicate NFW mass profiles fitted to the medians of MANGA, \HI and $(r_\mathrm{h}, M_\mathrm{h})$ data.
See \sref{sec:method} for details.
The results indicate a mass-dependent reduction of inner dark matter in observed massive halos.
}
\label{fig:popprofmhbins}
\end{figure*}

Using the large sample of $M_\mathrm{h}$ data 
along with the MaNGA and \HI data,
we 
study the mass profiles of central galaxies in various $M_\mathrm{h}$ bins. The profiles of both simulated and observed dynamical masses, $M_\mathrm{dyn}(<r)$, are shown in the left panels of Fig.~\ref{fig:popprofmhbins}. The IllustrisTNG (hereafter TNG) profiles (blue lines) show evident upturns at smaller radii, making the enclosed masses larger than the NFW profiles (black dashed lines; the NFW parameters are fitted in dark-matter-only simulations \citep{Ishiyama:2021MNRAS.506.4210I}) especially for larger $M_\mathrm{h}$ bins. This is consistent with the increasing dominance  of baryonic matter in the center of dark matter halos \citep{Schaller:2015MNRAS.451.1247S}. The TNG profiles are 
in general similar to
those in the EAGLE simulation, although the two simulations use different recipes for sub-grid physics
(see e.g.~\citet{Habouzit:2021MNRAS.503.1940H,Ward:2022MNRAS.514.2936W}, and \citet{Crain:2023ARA&A..61..473C} for a recent review). 

We next compare the observed mass profiles with the predictions of simulations. 
For low-mass halos, the observed dynamical masses are broadly consistent with the simulations from the inner galaxy to the \HI radius. In more massive halos, however, the \HI dynamical masses, $M_\mathrm{dyn}(<r_\mathrm{HI})$, become progressively lower than predicted, while the MaNGA measurements at the half-light radius, $M_\mathrm{dyn}(<r_\mathrm{1/2})$, remain broadly compatible with the simulated inner profiles. 
This offset is not explained by the changing mix of galaxy populations at higher halo mass: when the \HI sample is divided by star-formation activity and by visual morphology, the subsamples show differences in \HI mass and hence in $r_\mathrm{HI}$, but no significant systematic differences in rotational velocity $v_\mathrm{rot}$ or 
the offset relative to simulations
(see \sref{sec:method} and Extended Data Fig.~\ref{fig:hiprofselsfrmorph}).

The dark matter profiles obtained by subtracting the baryonic contribution, $M_\mathrm{DM}(<r)$, are shown in the right column of Fig.~\ref{fig:popprofmhbins}. The enclosed DM masses of TNG are slightly larger than the NFW profiles fitted from dark-matter-only simulations, which is consistent to TNG's prediction of enhanced dark matter masses within $\sim r_\mathrm{1/2}$ and contraction of halos, as discussed in relevant works \citep{Lovell:2018MNRAS.481.1950L}. The dark matter profiles of EAGLE are also consistent with those of TNG. 
The observationally inferred dark matter profiles
still remain approximately NFW-like (see pink dash-dotted lines for reference), but in massive halos they favor systematically lower enclosed dark-matter masses and lower effective concentrations than the simulations. 
The discrepancy therefore does not primarily indicate a breakdown of the NFW form itself; instead, it reflects a halo-mass-dependent deficit of inner dark matter at fixed total halo mass.

\begin{figure*}
\centering
\includegraphics[width=\linewidth]{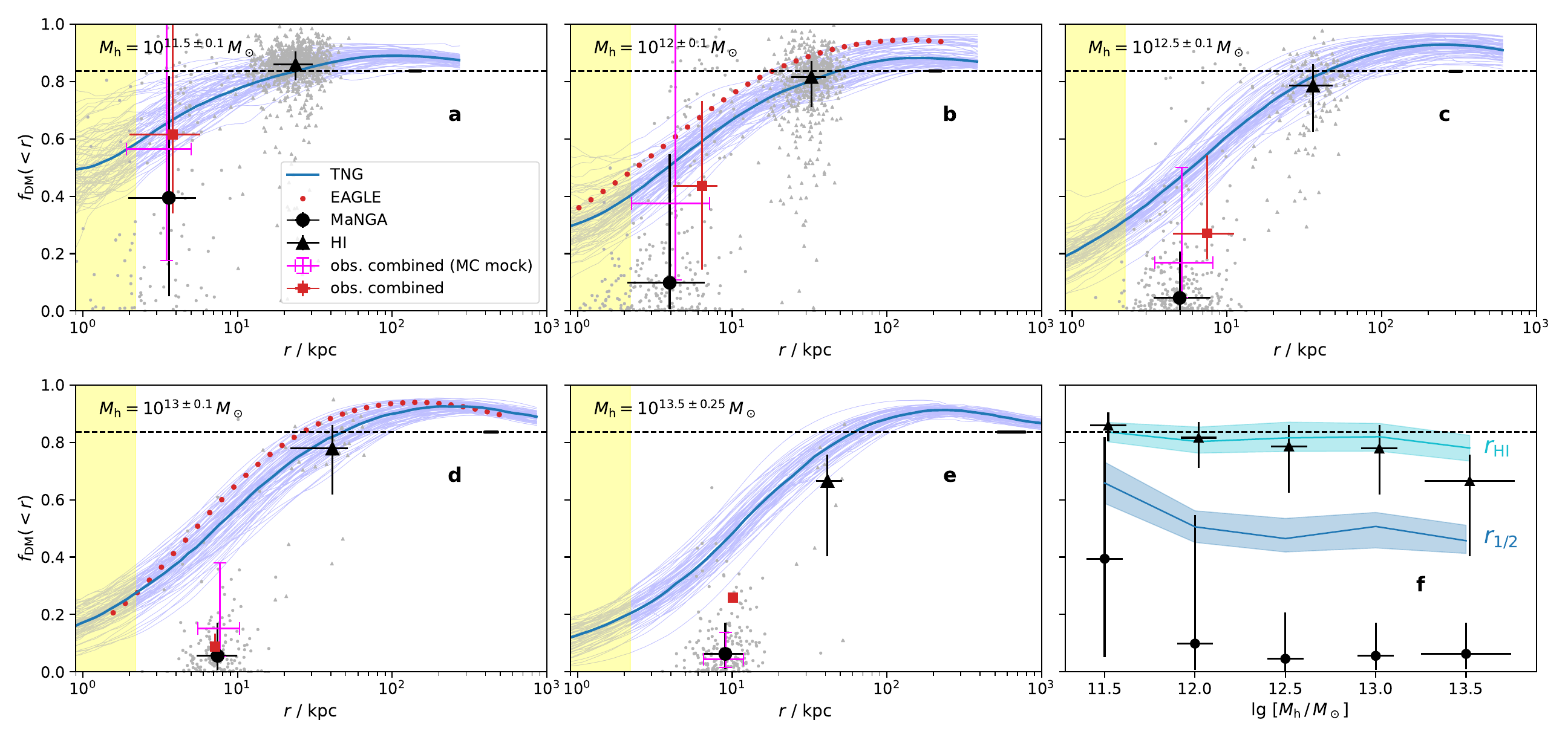}
\caption{\textbf{The dark matter fractions $f_\mathrm{DM}(<r)\equiv M_\mathrm{DM}(<r) / M_\mathrm{dyn}(<r)$.} 
\textit{Panels a--e}: similar to Fig.~\ref{fig:popprofmhbins}, each panel shows a specific $M_\mathrm{h}$ bin, with the same markers for both observational data (MaNGA: dots; \HI: triangles) and simulation data (TNG: blue lines, with yellow backgrounds indicating $r\lesssim 2.2\, \mathrm{kpc}$; EAGLE: red dots). 
The horizontal black dashed lines denote the cosmic value $f_\mathrm{DM} = 0.837$, equivalent to a cosmic baryon fraction of $f_\mathrm{bar} = 0.163$. 
The thick segments on the black dashed lines indicate the range of $r_\mathrm{h}$ within the $M_\mathrm{h}$ bins. 
The red squares and magenta error bars represent the estimation of $f_\mathrm{DM}$ at $r_\mathrm{1/2}$ based on the assumption that dark matter follows an NFW profile derived from $M_\mathrm{DM}(<r)$ for the $M_\mathrm{h}$ and \HI datasets, while $M_\mathrm{dyn}$ is measured by MaNGA stellar kinematics. This method is thus independent of the DM--star decomposition 
in dynamical modelings.
The red squares indicate median values for objects with both MaNGA and ALFALFA observations, with the 1-sigma scatters shown, except for the lower two panels (d, e) where only one or two galaxies are present in the bins. 
The magenta error bars indicate results where \HI data 
are assigned to MaNGA galaxies without ALFALFA observations by randomly selecting an \HI data point from the corresponding $M_\mathrm{h}$ bin,
which may overestimate the scatter of $f_\mathrm{DM}$ (see \sref{sec:method} for details). 
\textit{Panel f}: The median observational $f_\mathrm{DM}(<r)$ for MaNGA (dots, at $r_\mathrm{1/2}$) and \HI (triangles, at $r_\mathrm{HI}$) are shown as in the previous panels, with horizontal errorbars showing the bin sizes of $M_\mathrm{h}$. The TNG $f_\mathrm{DM}(<r)$ values at corresponding median radii are indicated by the blue lines (median) and shaded regions (1-sigma scatter). The results consistently suggest that the observed galaxies residing in large halos have systematically lower central dark matter fractions compared to simulations.
}
\label{fig:popproffdm}
\end{figure*}

This trend is seen even more clearly in 
the profiles of enclosed dark matter fractions, $f_\mathrm{DM}(<r)\equiv M_\mathrm{DM}(<r) / M_\mathrm{dyn}(<r)$, 
as shown in Fig.~\ref{fig:popproffdm}.
Unlike many previous studies,
here we plot $f_\mathrm{DM}(<r)$ as a function of physical radius $r$ across different $M_\mathrm{h}$ bins. This illustrates the distribution of baryons and dark matter, while avoiding inconsistencies 
arising from varying definitions of the effective radius $R_\mathrm{e}$ when comparing $f_\mathrm{DM}(<R_\mathrm{e})$.
For smaller aperture radii $r$, both simulations and observations show an increasing trend in $f_\mathrm{DM}$ with larger $r$, indicating the dominance of baryons in the central regions, which becomes more pronounced in larger halos. 
The TNG and EAGLE profiles of $f_\mathrm{DM}(<r)$ generally get close to the cosmic value at radii of order $r_\mathrm{HI}$, which is generally consistent with the observational results using \HI, indicating broad consistency at intermediate radii.

The discrepancy emerges in the central regions probed by MaNGA. 
Two independent observational estimates---one based on the MaNGA dynamical decomposition (black dots) and the other 
combining 
MaNGA,
\HI and $M_\mathrm{h}$ information
under an NFW assumption (red squares; see \sref{sec:method} for details)---both show that the central dark matter fraction, $f_\mathrm{DM}(<r_{1/2})$, decreases with halo mass more steeply than in TNG and EAGLE.
For low-mass halos (e.g.~$M_\mathrm{h}\sim 10^{11.5}M_\odot$), the observed central $f_\mathrm{DM}$ values are generally consistent with simulations within the population scatter, whereas
for massive halos they become systematically lower.
In larger $M_\mathrm{h}$ bins, the observed and simulated distributions of $f_\mathrm{DM}(<r_{1/2})$ are separated by about 4$\sigma$ (in units of the 1-sigma ranges of population scatters). 
The results therefore point to a mass-dependent reduction of inner dark matter in observed massive halos.

A natural interpretation of this mass-dependent inner dark-matter deficit is that massive halos have experienced stronger long-term baryonic halo heating, and hence weaker net contraction, than is captured by current hydrodynamical simulations. In the massive-halo regime, such heating may be driven not only by feedback from active galactic nuclei, but also by earlier merger- or satellite-induced dynamical-friction heating, which can pre-condition the inner halo and enhance the response to later gas outflows \citep{Dekel:2021MNRAS.508..999D}. The present-day mass profiles may therefore preserve the cumulative imprint of repeated baryonic perturbations over cosmic time. 
Importantly, our main result is robust to several potential systematics: differences in star-formation status or morphology do not produce significant shifts in $v_\mathrm{rot}$, restricting the \HI analysis to nearly edge-on galaxies does not change the inferred profiles, and adopting a different assumption regarding the initial mass function (IMF) does not significantly alter the trend in central dark-matter fraction (see \sref{sec:method} and Extended Data). Possible spatial offsets between the true halo center and the adopted central galaxy could, in principle, bias galaxy-centered estimates of the inner dark-matter fraction toward lower values, especially at small radii; however, conservative trimming tests indicate that such an effect is unlikely to account for the full deficit observed in massive halos (see \sref{sec:method}). This mechanism should therefore be regarded as a plausible physical interpretation rather than a demonstrated one. Future and ongoing \HI surveys, including the FAST All Sky \HI survey \citep{Zhang:2024SCPMA..6719511Z} and observations with the Square Kilometer Array \citep{Weltman:2020PASA...37....2W,Schilizzi:2024ska..book.....S}, will provide larger and deeper samples of both integrated and spatially resolved \HI data, enabling decisive tests of this result.

\section*{Methods}
\label{sec:method}

In this paper, we assume a flat \lcdm cosmology with $H_0 = 70\,\mathrm{km/s/Mpc}$, $\Omega_\mathrm{m} = 0.3$, $\Omega_\Lambda = 0.7$ and $\Omega_\mathrm{b} = 0.049$ wherever possible. In case an adopted dataset makes different assumptions, we make conversions \citep{Croton:2013PASA...30...52C} between different values of $h$. The halo mass $M_\mathrm{h}$ is defined as the virial mass $M_\mathrm{200c}$, the mass enclosed by a sphere of $r_\mathrm{200c}$ in which the mean density is $200$ times the cosmic critical density, $\rho_c$. All quantities in this paper are given in physical units (without factors of $h$).

\subsection*{Observational data}
\textit{MaNGA}. The Mapping Nearby Galaxies at Apache Point Observatory (MaNGA) \citep{Bundy:2015ApJ...798....7B} is an integral field spectroscopic survey that obtained spatially resolved spectra for $\sim 10\mathrm{k}$ galaxies. Using the 17 simultaneous fiber-bundle integral field units (IFUs) that feed into the Baryon Oscillation Spectroscopic Survey (BOSS) spectrographs \citep{Smee:2013AJ....146...32S,Drory:2015AJ....149...77D} of the Sloan $2.5\,\mathrm{m}$ Telescope \citep{Gunn:2006AJ....131.2332G}, spectral measurements were made across the face of galaxies. 
The radial coverage of MaNGA is out to 1.5  half-light radius $R_\mathrm{e}$ for the primary sample, and $2.5 R_\mathrm{e}$ for the secondary sample. By fitting the absorption lines of IFU spectra using the pPXF software \citep{Cappellari:2017MNRAS.466..798C} in the MaNGA Data Analysis Pipeline (DAP) \citep{Belfiore:2019AJ....158..160B,Westfall:2019AJ....158..231W}, stellar kinematic maps are extracted from the data cubes, with which dynamical models can be constructed to infer the mass distribution of galaxies.

We use the results of dynamical modeling from the MaNGA DynPop project \citep{Zhu:2023MNRAS.522.6326Z}, where stellar kinematics were fitted using the Jeans Anisotropic Modelling (JAM) method \citep{Cappellari:2008MNRAS.390...71C,Cappellari:2020MNRAS.494.4819C}. A total of eight models were fit to the observed root-mean-square velocity, $v_\mathrm{rms}$, each based on distinct assumptions. 
Each model assumes either a cylindrically-aligned (JAM\textsubscript{cyl}) or a spherically-aligned velocity ellipsoid (JAM\textsubscript{sph}), and a specific mass model that includes an extended mass distribution (stellar and dark matter) and a supermassive black hole. Given the best-fit model parameters, one can calculate the enclosed dynamical mass $M_\mathrm{dyn}(<r)$, as well as the decomposed contributions of dark matter and stellar mass. In this study, we adopt the results from their JAM\textsubscript{cyl}+gNFW model, 
where the extended mass includes
a generalized Navarro-Frenk-White (gNFW) \citep{Navarro:1996ApJ...462..563N,Zhao:1996MNRAS.278..488Z} dark matter halo, and the
deprojected stellar luminosity scaled by a spatially constant stellar mass-to-light ratio $M/L$.
We use the dynamical mass $M_\mathrm{dyn}(<r_{1/2})$ and dark matter mass $M_\mathrm{DM}(<r_{1/2})$ at the radius of $r_{1/2}$, which is the 3D half light radius derived from the deprojected luminosity profile \citep{Zhu:2023MNRAS.522.6326Z}. 

MaNGA DynPop also provides visual classifications of the fitting quality. With reference to their guidance \citep{Zhu:2023MNRAS.522.6326Z}, we require the fitting quality flag $\mathtt{Qual}\geq 1$ and the quality by MaNGA DRP pipeline $\mathtt{drp3qual} = 1$ (i.e., high quality). For $M_\mathrm{DM}(<r_{1/2})$, we additionally require that the JAM\textsubscript{sph} and JAM\textsubscript{cyl} models provide consistent results, $\left|f_\mathrm{DM}(<R_\mathrm{e}; \mathrm{sph}) - f_\mathrm{DM}(<R_\mathrm{e}; \mathrm{cyl})\right| < 0.1$.

\textit{ALFALFA}. The Arecibo Legacy Fast ALFA (ALFALFA) survey \citep{Giovanelli:2005AJ....130.2598G,Haynes:2011AJ....142..170H, Haynes:2018ApJ...861...49H} is a blind survey of the \HI 21\,cm line, mapping nearly 7000\,deg\textsuperscript{2} of the high Galactic latitude sky. The completed survey detects $\sim$31,500 extragalactic \HI line sources out to $z<0.06$. The \HI mass, $M_\mathrm{HI}$, can be calculated from integrated \HI line flux, $F$, and distance, $d$, via the standard formula \citep{Roberts:1962AJ.....67..437R}, 
\begin{gather}
\frac{M_{\mathrm{HI}}}{M_\odot}=2.356 \times 10^5 \left(\frac{d}{\mathrm{Mpc}}\right)^2 \frac{F}{\mathrm{Jy\,km\,s^{-1}}}.\end{gather}
The widths of \HI lines can be a proxy of rotational velocities, $v_\mathrm{rot}$. We adopt the determination of $v_\mathrm{rot}$ in \citet{Yu:2022ApJS..261...21Y}, where line widths are measured using the curve of growth method, corrected for inclination angle, instrumental and turbulent broadening, and calibrated with $v_\mathrm{rot}$ from spatially resolved observations \citep{Yu:2020ApJ...898..102Y}. With $v_\mathrm{rot}$, the dynamical mass at the \HI radius, $r_\mathrm{HI}$, defined at a surface density of $1\,M_\odot\,\mathrm{pc}^{-2}$, is estimated by 
\begin{gather}
M_\mathrm{dyn} (< r_\mathrm{HI}) = \frac{v_\mathrm{rot}^2 r_\mathrm{HI}}{G},
\label{eq:Mdyn-vrot}
\end{gather}
where $r_\mathrm{HI}$ is calculated using the tight \HI mass--size relation \citep{Wang:2016MNRAS.460.2143W},
\begin{gather}
\lg r_{\mathrm{HI}} = 0.506 \lg M_{\mathrm{HI}} - 3.293  - \lg 2,
\end{gather}
where $\lg\equiv\log_{10}$.

The ALFALFA catalog provides quality codes of \HI detection. Code 1 refers to sources of highest quality, while code 2 refers to sources of low S/N but have optical counterparts with known optical redshifts coincident with those of the \HI lines (see \citet{Haynes:2018ApJ...861...49H} for details). \citet{Yu:2022ApJS..261...21Y} also provides flags of sources that are affected by a low fraction of outlying bad channels, or potentially blended with nearby companions. To avoid any possible factors that increase the errors of the results, we only use galaxies with quality code 1 and without any flags of bad channels or possible companions.

\textit{Halo masses of SDSS groups}. 
One traditional method for statistically estimating halo masses is abundance matching
\citep[e.g.][]{Frenk:1988ApJ...327..507F,Kravtsov:2004ApJ...609...35K,Tasitsiomi:2004ApJ...614..533T,Yang:2007ApJ...671..153Y}, 
which assumes that the most massive (or luminous) galaxy group occupies the most massive dark matter halo, the second most massive group
resides 
in the second most massive halo, 
etc.
However, 
such a simple relation of galaxies and halos can introduce biases by neglecting the dependence of halo mass on galaxy color \citep{Mandelbaum:2016MNRAS.457.3200M,Bilicki:2021A&A...653A..82B,Man:2019ApJ...881...74M}, halo assembly history (or, observationally, star formation history) \citep{Matthee:2017MNRAS.465.2381M,Lyu:2023ApJ...959....5L,Lyu:2024ApJ...972..108L}, etc. 

In this work, we use the halo mass estimations of galaxy groups in SDSS DR7 obtained from a machine learning model trained on mock galaxy catalogs \citep{Zhao:2025ApJ...979...42Z}.
As demonstrated by \citet{Zhao:2025ApJ...979...42Z}, 
the halo mass results
are
consistent with the theoretical halo mass function and the stellar-to-halo mass relations for blue and red groups directly measured through weak lensing. Compared to traditional abundance matching results, the $M_\mathrm{h}$ estimation from the machine learning model shows a remarkably better alignment with weak lensing observations, highlighting the improvement of this approach.

When generating the $M_\mathrm{h}$ dataset, the group catalog was constructed from the NYU-VAGC DR7 catalog \citep{Blanton:2005AJ....129.2562B} using a refined adaptive halo-based group finder \citep{Yang:2005MNRAS.356.1293Y,Yang:2007ApJ...671..153Y}. The stellar masses, star formation rates (SFRs), stellar ages, colors, etc.\ of galaxies in the groups were used as observational parameters, with which halo masses were inferred using an \lstinline|XGBoost| \citep{Chen2016XGBoost:10.1145/2939672.2939785} model. The model was trained on a mock galaxy catalog generated from the L-GALAXIES semi-analytical model \citep{Henriques:2015MNRAS.451.2663H}. To ensure the mock data match the SDSS observations and sample selections, observational uncertainties (simulated by introducing Gaussian noise) and data normalization were 
considered.

We use the $M_\mathrm{h}$ predictions from the model trained on the mock dataset with a noise level that matches the observational noise. 
Since the output $M_\mathrm{h}$ data are total masses of galaxy groups rather than subhalo masses of satellite galaxies, we only study central galaxies in this work. 
This also mitigates potential environmental effects, 
which can be effective when studying satellite galaxies.

\textit{Data compilation}. 
We match the SDSS $M_\mathrm{h}$ catalog \citep{Zhao:2025ApJ...979...42Z} with the aforementioned MaNGA and ALFALFA catalogs to obtain the final sample.
For the ALFALFA \HI catalog, we merge the SDSS subsample in \citet{Yu:2022ApJS..261...21Y} with the $M_\mathrm{h}$ catalog using a maximum optical position separation of $3''$. The MaNGA DynPop catalog \citep{Zhu:2023MNRAS.522.6326Z} is matched according to the NASA-Sloan Atlas\footnote{\url{http://nsatlas.org/}} (NSA) catalog \citep{Blanton:2011AJ....142...31B}. We match the stellar mass $M_\star$ and star formation rate (SFR) from the MPA-JHU catalog \citep{Kauffmann:2003MNRAS.341...33K,Brinchmann:2004MNRAS.351.1151B,Tremonti:2004ApJ...613..898T}, and adopt their median values. 
Finally, we obtain a sample of 10,438 central galaxies, with 5,608 matched to ALFALFA \HI measurements and 5,256 matched to the MaNGA DynPop catalog.

\subsection*{Simulation data}
For comparison to observations, we mainly use the simulated galaxies from the
IllustrisTNG project\footnote{\url{https://www.tng-project.org/}} \citep{Marinacci:2018MNRAS.480.5113M,Naiman:2018MNRAS.477.1206N,Nelson:2018MNRAS.475..624N,Nelson:2019ComAC...6....2N,Nelson:2019MNRAS.490.3234N,Pillepich:2018MNRAS.475..648P,Pillepich:2019MNRAS.490.3196P,Springel:2018MNRAS.475..676S}. As a followup to the Illustris project \citep{Vogelsberger:2014Natur.509..177V,Vogelsberger:2014MNRAS.444.1518V}, TNG simulates the evolution of halos and galaxies using the \lstinline|AREPO| moving-mesh code \citep{Springel:2010MNRAS.401..791S}. 
The original TNG project consists of three simulation volumes, namely TNG50, TNG100, and TNG300, 
where the numbers indicate the physical side lengths of simulation boxes. 
We adopt the TNG100 simulation, which has a cubic volume of roughly 100 Mpc side length, and a resolution of $1.4\times 10^6M_\odot$ and $7.5\times 10^6 M_\odot$ for baryon and dark matter, respectively.

We extract the mass profiles of central galaxies from TNG100. In the simulation, dark matter halos and subhalos are identified using the friends-of-friends (FoF) algorithm \citep{Davis:1985ApJ...292..371D} and Subfind algorithm \citep{Springel:2001MNRAS.328..726S,Dolag:2009MNRAS.399..497D}, respectively. Baryonic substructures corresponding to the dark matter subhalos are designated as galaxies, and a central galaxy/subhalo is typically the one with the minimal gravitational potential. 
To extract the mass profiles, we construct 100 logarithmically spaced radial bins from 0.03 kpc to 3000 kpc and sum the mass of all dark matter, gas, and star particles enclosed within each radius separately. Since the enclosed mass can include satellite galaxies at large radii, we only plot the profiles within $\sim 2 r_\mathrm{h}$. 

In addition to TNG, we also extract the mass profiles of stacked halos in the EAGLE simulation. The results are reproduced from Fig.\ 6 in \citet{Schaller:2015MNRAS.451.1247S}, where halos were stacked by coadding halos in bins of width $\Delta \lg M_\mathrm{h} = 0.2$.

\subsection*{Estimation of masses, profiles, and dark matter fractions}
\textit{Enclosed masses}. 
For TNG and EAGLE simulations,
the enclosed dynamical mass profiles
are calculated by summing the profiles of dark matter, star, and gas components. For the MaNGA DynPop catalog, we primarily use the decomposition of dark matter and baryon (stellar) mass from their dynamical models, as described above. For the \HI and $M_\mathrm{h}$ data, since only the total dynamical mass is provided, we estimate the enclosed dark matter mass by subtracting the baryon mass from the dynamical mass. The baryon mass within $r_\mathrm{HI}$ is estimated by 
\begin{gather}
M_\mathrm{bar} = M_\star + 1.33 M_\mathrm{HI},
\end{gather}
where $1.33 M_\mathrm{HI}$ is the gas mass considering the contribution of Helium. Although the molecular gas is neglected here, its contribution to the total baryon mass of nearby galaxies is generally small \citep[e.g.][]{Cortese:2014ApJ...795L..37C,McGaugh:2015ApJ...802...18M,Lelli:2019MNRAS.484.3267L}. 
Finally, we estimate the enclosed baryon mass at $r_\mathrm{h}$ by $f_\mathrm{bar}M_\mathrm{h}$, where $f_\mathrm{bar}\approx 0.16$ is the cosmic baryon fraction.

\textit{The NFW profile}. It was shown that the spherically averaged mass distribution of relaxed halos in $N$-body (cold dark matter) simulations satisfies the Navarro–Frenk–White (NFW) profile \citep{Navarro:1996ApJ...462..563N,Navarro:1997ApJ...490..493N}. The NFW density profile is defined as:
\begin{gather}
\rho^\mathrm{NFW}(r; r_\mathrm{s}, \rho_\mathrm{s})=\frac{4 \rho_{\mathrm{s}}}{\left(r / r_{\mathrm{s}}\right)\left(1+r / r_{\mathrm{s}}\right)^2},
\end{gather}
where $r_\mathrm{s}$ is the scale radius and $\rho_\mathrm{s}$ is the density at $r_\mathrm{s}$.
The NFW profile can also be expressed using virial mass $M_\Delta$ and concentration $c_\Delta$ as free parameters, where concentration is defined as $c_\Delta\equiv r_\Delta/r_\mathrm{s}$. Here, $r_\Delta$ is the virial radius, defined at the radius where the mean enclosed density (i.e., the enclosed mass $M_\Delta$ divided by the enclosed volume of the sphere) is $\Delta\rho_\mathrm{c}$, where $\rho_\mathrm{c}={3H^2}/{8\pi G}$ is the cosmic critical density. 
The mass enclosed within a radius $r$ for an NFW profile is
\citep[e.g.][]{Bhattacharya:2013ApJ...766...32B,Freundlich:2020MNRAS.499.2912F}
\begin{gather}
M^\mathrm{NFW}(<r; M_\Delta, c_\Delta) = \frac{m(c_\Delta r/r_\Delta)}{m(c_\Delta)} M_\Delta,
\end{gather}
where 
\begin{gather}
m(y) \equiv \ln (1+y) - \frac{y}{1+y}.
\end{gather}
As we take $\Delta=200$ throughout this paper, we denote halo mass (virial mass) $M_\mathrm{h}\equiv M_\mathrm{200c}$, concentration $c_\mathrm{h}\equiv c_\mathrm{200c}$, and virial radius $r_\mathrm{h}\equiv r_\mathrm{200c}$, for simplicity. 

The relation between concentration $c_\mathrm{h}$ and halo mass $M_\mathrm{h}$ has been extensively studied in $N$-body simulations \citep[e.g.][]{Bullock:2001MNRAS.321..559B,Duffy:2008MNRAS.390L..64D,Prada:2012MNRAS.423.3018P,Bhattacharya:2013ApJ...766...32B,Dutton:2014MNRAS.441.3359D,Klypin:2016MNRAS.457.4340K,Ludlow:2016MNRAS.460.1214L,Child:2018ApJ...859...55C,Diemer:2019ApJ...871..168D,Wang:2020Natur.585...39W,Ishiyama:2021MNRAS.506.4210I}.
In Fig.\ \ref{fig:popprofmhbins}, we plot NFW profiles (black dashed lines) for various bins of $M_\mathrm{h}$ using the $M_\mathrm{h}$--$c_\mathrm{h}$ relation of the relaxed halos at redshift $z=0$ in the Uchuu $N$-body simulations \citep{Ishiyama:2021MNRAS.506.4210I}. In the mass range of interest in our study, concentration decreases with increasing halo mass. 
We note that the $M_\mathrm{h}$--$c_\mathrm{h}$ relation 
can vary slightly across different studies,
depending on the definition of concentration, the selection of the dynamical state of halos, etc.
Nevertheless,
we do not expect the differences to significantly impact our results.

\textit{Alternative estimations of halo mass from \HI}. We use halo mass estimations in \citet{Zhao:2025ApJ...979...42Z} as the fiducial value. For comparison, we also estimate $M_\mathrm{h}$ using only the \HI radius and dynamical mass data with two different approaches. 
In the first approach, we calculate $M_\mathrm{h}$ from an empirical relation of $M_\mathrm{h}$ and the \HI dynamical mass, $M_\mathrm{dyn}(<r_\mathrm{HI})$, as in \citet{Yu:2020ApJ...898..102Y}:
\begin{gather}
\lg M_\mathrm{h} = 0.90 \lg M_\mathrm{dyn}(<r_\mathrm{HI}) + 1.76.
\end{gather}
In the second approach, we use not only $M_\mathrm{dyn}(<r_\mathrm{HI})$ but also the radius $r_\mathrm{HI}$ to calculate $M_\mathrm{h}$. 
Here we assume an NFW profile going through the \HI data,
\begin{gather}
M_\mathrm{dyn}(<r_\mathrm{HI}) = M^\mathrm{NFW}(<r_\mathrm{HI}; M_\mathrm{h}, c_\mathrm{h}),
\label{eq:1ptfit}
\end{gather}
and the parameters $M_\mathrm{h}$ and $c_\mathrm{h}$ are related by
the following relation \citep{Dutton:2014MNRAS.441.3359D}:
\begin{gather}
\lg c_\mathrm{h}=0.905-0.101 \lg\left(M_\mathrm{h} /\left[10^{12} h^{-1} \mathrm{M}_{\odot}\right]\right),
\label{eq:McDutton}
\end{gather}
where $h=0.7$. Combining Eqs.\ (\ref{eq:1ptfit}) and (\ref{eq:McDutton}), we can obtain the solutions of $M_\mathrm{h}$ and $c_\mathrm{h}$. The results of the two approaches are shown in the two panels of Fig.~\ref{fig:comphihalomass}.

\textit{Dark matter fractions}. Given the enclosed dark matter mass, $M_\mathrm{DM}(<r) $, and dynamical mass, $M_\mathrm{dyn}(<r)$, the  enclosed dark matter fraction is calculated by
\begin{gather}
f_\mathrm{DM}(<r)\equiv \frac{M_\mathrm{DM}(<r)}{M_\mathrm{dyn}(<r)},
\end{gather}
for both observational and simulation data.

Since the dark matter fraction 
from MaNGA,
$f_\mathrm{DM}(<r_{1/2})$, can be affected by the mass decomposition in the JAM dynamical modeling, we also make an alternative estimation that is independent of that decomposition. 
We assume an NFW profile for the dark matter component, $M_\mathrm{DM}(<r)$, and
fit the NFW profile to the data estimated from \HI and $M_\mathrm{h}$, i.e., $M_{\mathrm{DM}}(<r_\mathrm{HI})$ and $M_{\mathrm{DM}}(<r_\mathrm{h})$.
We then combine the derived dark matter component with the $M_\mathrm{dyn}(<r_{1/2})$ from the MaNGA DynPop catalog to obtain an alternative estimation of $f_\mathrm{DM}(<r_{1/2})$. In Fig.\ \ref{fig:popproffdm}, the red squares indicate the results using only the sources with both MaNGA and ALFLAFA observations, so that we can obtain corresponding values of $M_{\mathrm{DM}}(<r_\mathrm{HI})$, $M_{\mathrm{DM}}(<r_\mathrm{h})$, and $M_\mathrm{dyn}(<r_{1/2})$. However, the sample size is small for larger $M_\mathrm{h}$ bins.
Therefore, we also make an estimation for each galaxy with $M_\mathrm{dyn}(<r_{1/2})$ from MaNGA, but without ALFALFA observation, by assigning it a $M_{\mathrm{DM}}(<r_\mathrm{HI})$ value randomly selected from the corresponding $M_\mathrm{h}$ bin. The results are marked in magenta in Fig.\ \ref{fig:popproffdm}. We note that this assignment neglects the correlation between MaNGA and ALFAFLA data, which can be partly due to the range of radii and masses within the bin size, and thus may overestimate the scatters. Nevertheless, both the results from dynamical decompositions and our alternative methods described above consistently show a systematically lower central $f_\mathrm{DM}$ for large $M_\mathrm{h}$ bins compared to simulations.

\subsection*{Discussion and tests of robustness}

The low dark matter fraction $f_\mathrm{DM}$ in the central regions of halos shown in this paper is consistent to previous studies using stellar kinematics and dynamical modelings \citep[e.g.][]{Cappellari:2013MNRAS.432.1862C,Zhu:2024MNRAS.527..706Z,Yang:2024MNRAS.528.5295Y}.
In addition to stellar kinematics, other tracers such as
strong gravitational lensing \citep[e.g.][]{Barnabe:2011MNRAS.415.2215B}, globular cluster and satellite kinematics \citep[e.g.][]{Alabi:2017MNRAS.468.3949A,Wojtak:2013MNRAS.428.2407W}  have also been used in previous studies to estimate $f_\mathrm{DM}$. 
The estimates of 
$f_\mathrm{DM}$ obtained using different approaches yield varying results \citep{Lovell:2018MNRAS.481.1950L},
suggesting that the choice of tracers and modeling assumptions can 
influence the outcomes.

In dynamical modeling, particularly the decomposition of dark matter and baryons within baryon-dominated regions, assumptions regarding the potential shape, conversion between light and mass profiles (i.e., the mass-to-light ratio), and orbital anisotropy may impact the output results. Because variations in the IMF drive systematic drifts in stellar mass, the degeneracy between stellar and dark matter components
may introduce uncertainties in the dark matter mass within $r_{1/2}$ \citep{Lovell:2018MNRAS.481.1950L}. 
For spatially integrated \HI observations, the derivation of rotational velocity and dynamical mass from line-width measurements and the asymmetric drift correction can be additional sources of potential error \citep{Ho:2007ApJ...668...94H}. Refining these mass derivation and decomposition can be an important direction for future work.

There could also be systematic errors originating from the sample selection, potentially making the sample 
not fully representative of the whole galaxy population. For example, \HI emission line observations may have a lower detection rate for smaller \HI masses (which decreases the total flux from the \HI lines) and for larger line widths (which distributes the flux across more frequency channels). The trend of \HI results for increasing $M_\mathrm{h}$ might also be interpreted as an increasing fraction of passive elliptical galaxies, which have smaller \HI masses. However, we find that the difference in the \HI mass among various morphologies or star formation rates cannot account for the systematic difference of the \HI dynamical mass (see Extended Data Fig.~\ref{fig:hiprofselsfrmorph}). Future and ongoing \HI surveys such as the FAST All Sky \HI (FASHI) survey \citep{Zhang:2024SCPMA..6719511Z} will 
deepen and enlarge
the \HI sample, 
allowing for more robust confirmation of the results.

We also note that the simulations are compared directly to observations, without generating mock observations or simulating real survey conditions. Future work involving detailed simulations of observational processes will enable a more 
rigorous
comparison between simulations and observations.

In the following, we report 
several tests on the robustness of our results, including the selection of star formation status, morphology, and variations in the initial mass function (IMF) applied to the $M_\star$ data. We find no significant impact from these aspects on our results.

\textit{Star formation status and morphology}. 
It is 
generally expected  that galaxies in larger halos 
are more dominated by passive, elliptical galaxies, which typically have smaller $M_\mathrm{HI}$ compared to their star-forming, late-type counterparts.
 To investigate whether differences in star formation status and morphology affect the mass profiles, we divide the galaxies into several classes based on their star formation rates (SFRs) and their visual morphological classifications from Galaxy Zoo 1 \citep{Lintott:2008MNRAS.389.1179L,Lintott:2011MNRAS.410..166L}. Following \citet{Trussler:2020MNRAS.491.5406T}, we divide galaxies into star forming (SF), green valley (GV), and passive (P) galaxies on the $M_\star$--SFR plane. Specifically, SF galaxies are defined by $\lg\mathrm{SFR} - 0.70 \lg M_\star > -7.52$, passive galaxies by $\lg\mathrm{SFR} - 0.70 \lg M_\star < -8.02$, and the remaining are defined as GV galaxies (as illustrated in Fig.\ 1 of \citet{Trussler:2020MNRAS.491.5406T}). For morphological classifications, we use the Galaxy Zoo 1 morphology flags: ``spiral'', ``elliptical'', and ``uncertain''. The ``spiral'' and ``elliptical'' classifications require an 80\% debiased vote form volunteers, while 
galaxies that do not meet this criterion 
are flagged as ``uncertain''.

The influence of star formation status and morphology on the dynamical mass probed by \HI, $M_\mathrm{dyn}(<r_\mathrm{HI})$, is shown in Extended Data Fig.\ \ref{fig:hiprofselsfrmorph}. As can be seen, spiral galaxies and star-forming galaxies generally have larger $M_\mathrm{HI}$, and thus larger $r_\mathrm{HI}$, in most $M_\mathrm{h}$ bins. However, despite the difference in radius, 
there are no
significant systematic differences in $v_\mathrm{rot}$.
By comparing the \HI data to the TNG simulation, which exhibit approximately constant $v_\mathrm{rot}$ near $r_\mathrm{HI}$, we find that 
any potential preference for
star formation status or morphology 
within specific $M_\mathrm{h}$ bins does not account for the systematic differences between observation and simulation.

\textit{Inclination angle}. In the \HI dataset, the rotational velocity $v_\mathrm{rot}$ is determined from the observed line width considering the inclination angle $i$ and broadening effects \citep{Yu:2020ApJ...898..102Y}. For nearly face-on galaxies, the small radial velocity results in a small line width, and potentially  a large relative error in the final $v_\mathrm{rot}$ result. We reproduce the mass profiles in Fig.\ \ref{fig:popprofmhbins} restricting to nearly edge-on galaxies ($i > 60^\circ$),
and do not find a significant difference in the results (Extended Data Fig.~\ref{fig:popprofmhbinsseli}). To obtain a reasonable sample size, we do not make selections on the inclination angle in our main results.

\textit{Variation in the initial mass function and $M_\star$}. For the MaNGA DynPop's
stellar and dark matter mass decomposition
(and thus its $f_\mathrm{DM}$ data), the stellar mass is dynamically determined without artificial assumptions on the initial mass function (IMF). As a result, the dynamically derived stellar mass to light ratio, $M_\star/L$, can be used to constrain the systematic variation of IMF \citep[e.g.][]{Cappellari:2012Natur.484..485C,Tortora:2013ApJ...765....8T,Mehrgan:2024ApJ...961..127M}.
On the other hand, our estimation of the enclosed dark matter mass at $r_\mathrm{HI}$, $M_\mathrm{DM}(<r_\mathrm{HI})$, is based on the MPA-JHU stellar mass $M_\star$ \citep{Kauffmann:2003MNRAS.341...33K}, which assumes a Kroupa IMF \citep{Kroupa:2001MNRAS.322..231K}. 
To test the impact of the variation in the IMF, we recalculate $f_\mathrm{DM}$ profiles (Fig.\ \ref{fig:popproffdm}) with the MPA-JHU $M_\star$ divided by $\alpha = 0.66$, which is approximately the $M_\star$ conversion \citep{Madau:2014ARA&A..52..415M} from Kroupa to Salpeter \citep{Salpeter:1955ApJ...121..161S} IMF. This results in an increase in $M_\star$ and an decrease in $M_\mathrm{DM}(<r_\mathrm{HI})$, and therefore a decrease in $f_\mathrm{DM}(<r_\mathrm{HI})$ as well as the alternative $f_\mathrm{DM}(<r_{1/2})$ estimation (mentioned above) combining the three datasets (see Extended Data Fig.\ \ref{fig:popproffdmimfhi}). Nevertheless, this variation in IMF does not significantly affect the conclusions in Fig.\ \ref{fig:popproffdm}.

\subsection*{Halo-center offsets (mis-centering)}

The most massive galaxy in a group or cluster is not always located at the true halo center. In X-ray-selected SDSS clusters, about 65\% of systems have their most massive member near the X-ray flux peak, while the remaining $\sim 35\%$ show non-zero offsets \citep{Skibba:2011MNRAS.410..417S,Wang:2014MNRAS.439..611W}. Such offsets can bias galaxy-centered estimates of the enclosed dark-matter fraction toward lower values, because the aperture is centered on the galaxy rather than on the minimum of the host-halo potential.

A simple toy model illustrates the scale of the effect. For an NFW halo with $M_\mathrm{h}=10^{13}M_\odot$, a representative concentration, and a central galaxy of $M_\star\sim10^{11}M_\odot$, an offset of $100\,\mathrm{kpc}$ lowers the inferred $f_\mathrm{DM}(<10\,\mathrm{kpc})$ from $\sim0.56$ to $\sim 0.015$, and $f_\mathrm{DM}(<40\,\mathrm{kpc})$ from $\sim 0.9$ to $\sim 0.43$, in a pure host-halo calculation.  These numbers show that mis-centering can, in principle, substantially reduce the inferred inner dark-matter fraction, especially for the smaller MaNGA aperture. At the same time, they should be regarded as an upper-limit style estimate, because the presence of a surviving subhalo around the offset galaxy would make the reduction less extreme.

To assess whether mis-centering can plausibly drive our main results, we perform a deliberately conservative trimming test in the massive halo bins. Motivated by the observational fraction above, we assume a scenario where 35\% of systems are significantly affected by mis-centering and remove the lowest 35\% of the observed MaNGA $f_\mathrm{DM}(<r_{1/2})$ values in each halo-mass bin. Even under this extreme assumption, the remaining upper 65\% of the observed distribution remains systematically below the TNG/EAGLE predictions in massive halos by approximately $3.5\sigma$ (for the $M_\mathrm{h}\approx 10^{13.5} M_\odot$ bin) to $4 \sigma$ (for the $M_\mathrm{h} \approx 10^{13} M_\odot$ bin), indicating that mis-centering is unlikely to be the dominant origin of the observed inner dark-matter deficit.

In addition, our primary conclusion is differential. In our TNG analysis, the practical definition of the central is matched to the observational one as closely as possible.  Therefore, the same type of effect should also be present, at least to some extent, in the simulated comparison. Since our main result is that observed massive halos contain less inner dark matter than predicted, we expect the differential conclusion to remain robust.

\bmhead{Acknowledgments}
We thank Niankun Yu, Kai Zhu for useful discussions. 
Y.P.\ and Y.-C.W.\ acknowledge support from the National SKA Program of China under grant No.\ 2025SKA0150102 and from the National Natural Science Foundation of China (NSFC) under grant Nos.\ 12125301, 12192220, and 12192222. Y.P.\ also acknowledges support from the New Cornerstone Science Foundation through the XPLORER PRIZE.
J.D.\ acknowledges the support of National Science Foundation of China (NSFC) grant Nos.\ 12303010.  This work is supported by the High-performance Computing Platform of Peking University.

\begin{appendices}

\begingroup
\renewcommand{\figurename}{Extended Data Fig.}

\begin{figure*}
\centering
\includegraphics[width=0.7\linewidth]{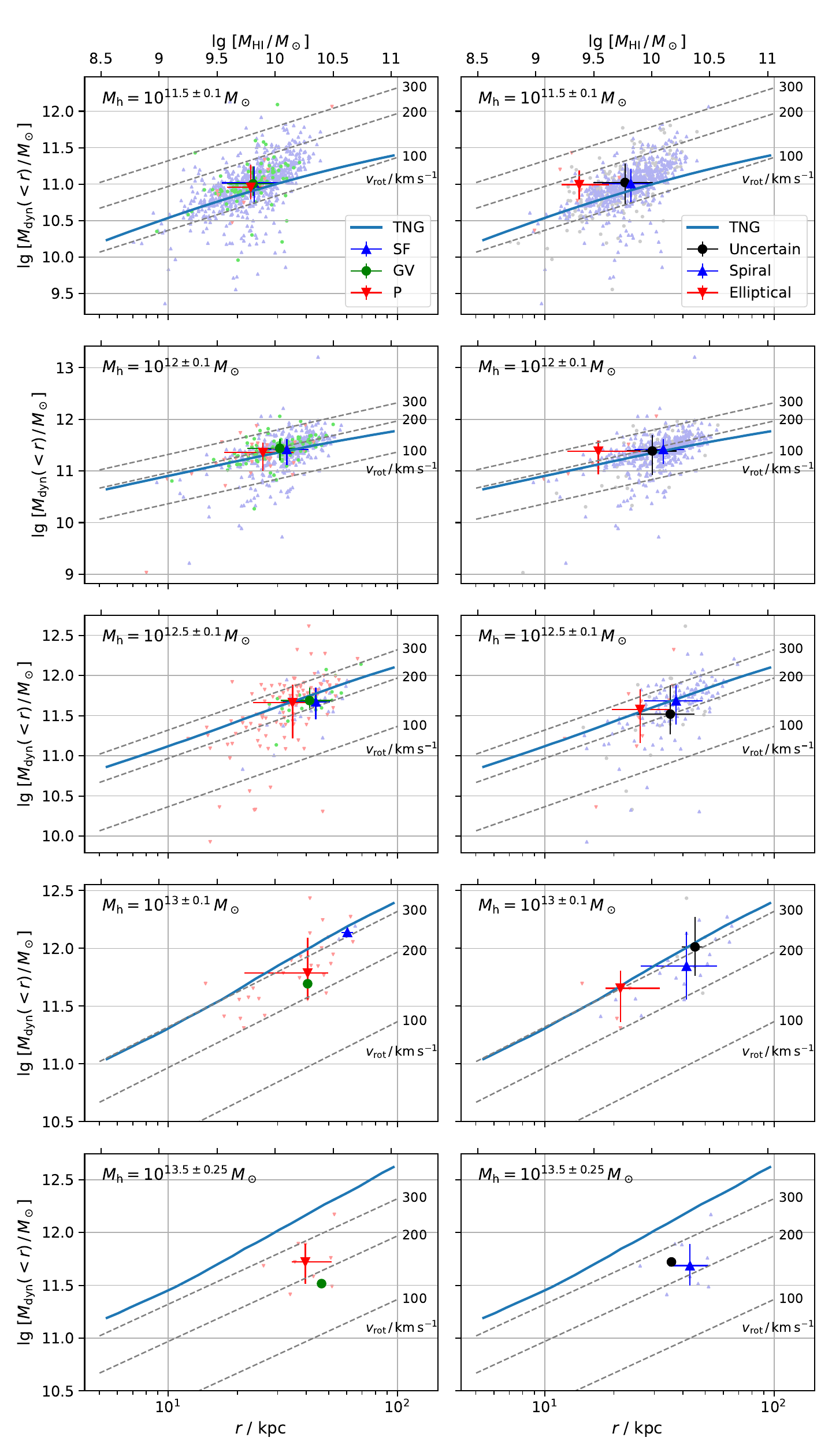}
\caption{\textbf{Comparison of} \HI \textbf{data $(r_\mathrm{HI}, M_\mathrm{dyn}(<r_\mathrm{HI}))$ for different star formation rates and morphologies.} Different rows show the same halo mass ($M_\mathrm{h}$) bins as in the text. 
The blue lines represent the median dynamical mass profiles
($M_\mathrm{dyn}(<r)$, the enclosed total mass within radius $r$)
in TNG. The \HI data, $r_\mathrm{HI}$ and $M_\mathrm{dyn}(<r_\mathrm{HI})$, are shown with small points (for individual data) and large points with errorbars (representing the medians and 1-sigma ranges), similar to Fig.~\ref{fig:popprofmhbins}. 
\textit{Left panels}: the \HI sample is divided into star-forming (SF, blue), green valley (GV, green), and passive (P, red) galaxies. 
\textit{Right panels}: the \HI sample is divided into three morphological classes: elliptical (red), spiral (blue), and uncertain (black).
Since the $r_\mathrm{HI}$ is calculated using the $M_\mathrm{HI}$--$r_\mathrm{HI}$ relation, the corresponding \HI masses ($M_\mathrm{HI}$) are also labeled on the $x$-axes. As related to $r_\mathrm{HI}$ and $M_\mathrm{dyn}(<r_\mathrm{HI})$ by Eq.~(\ref{eq:Mdyn-vrot}), the corresponding rotational velocities ($v_\mathrm{rot}$) are shown with the gray dashed lines. 
Although galaxies with different star formation rates or morphologies show differences in $M_\mathrm{HI}$ (and thus $r_\mathrm{HI}$) in some $M_\mathrm{h}$ bins, they have no significant systematic difference in $v_\mathrm{rot}$ and $M_\mathrm{dyn}$. 
} 
\label{fig:hiprofselsfrmorph}
\end{figure*}

\begin{figure*}
\centering
\includegraphics[width=.8\linewidth]{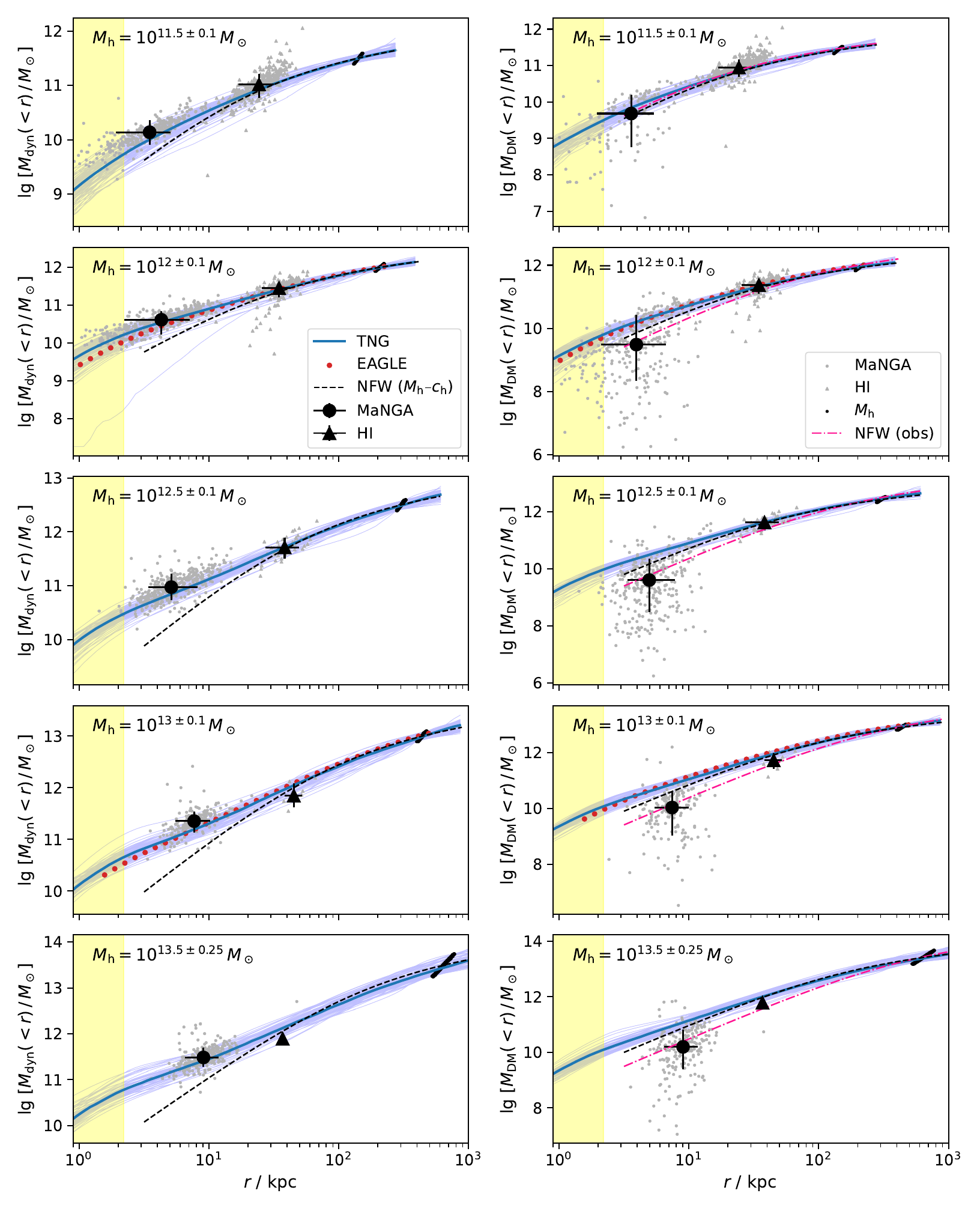}
\caption{\textbf{Dynamical mass $M_\mathrm{dyn}(<r)$ and dark matter mass $M_\mathrm{DM}(<r)$ profiles selecting nearly edge-on galaxies for }\HI \textbf{data.}  Same as Fig.~\ref{fig:popprofmhbins}, but the \HI data only include galaxies with inclination $i>60^\circ$. 
}
\label{fig:popprofmhbinsseli}
\end{figure*}

\begin{figure*}
\centering
\includegraphics[width=\linewidth]{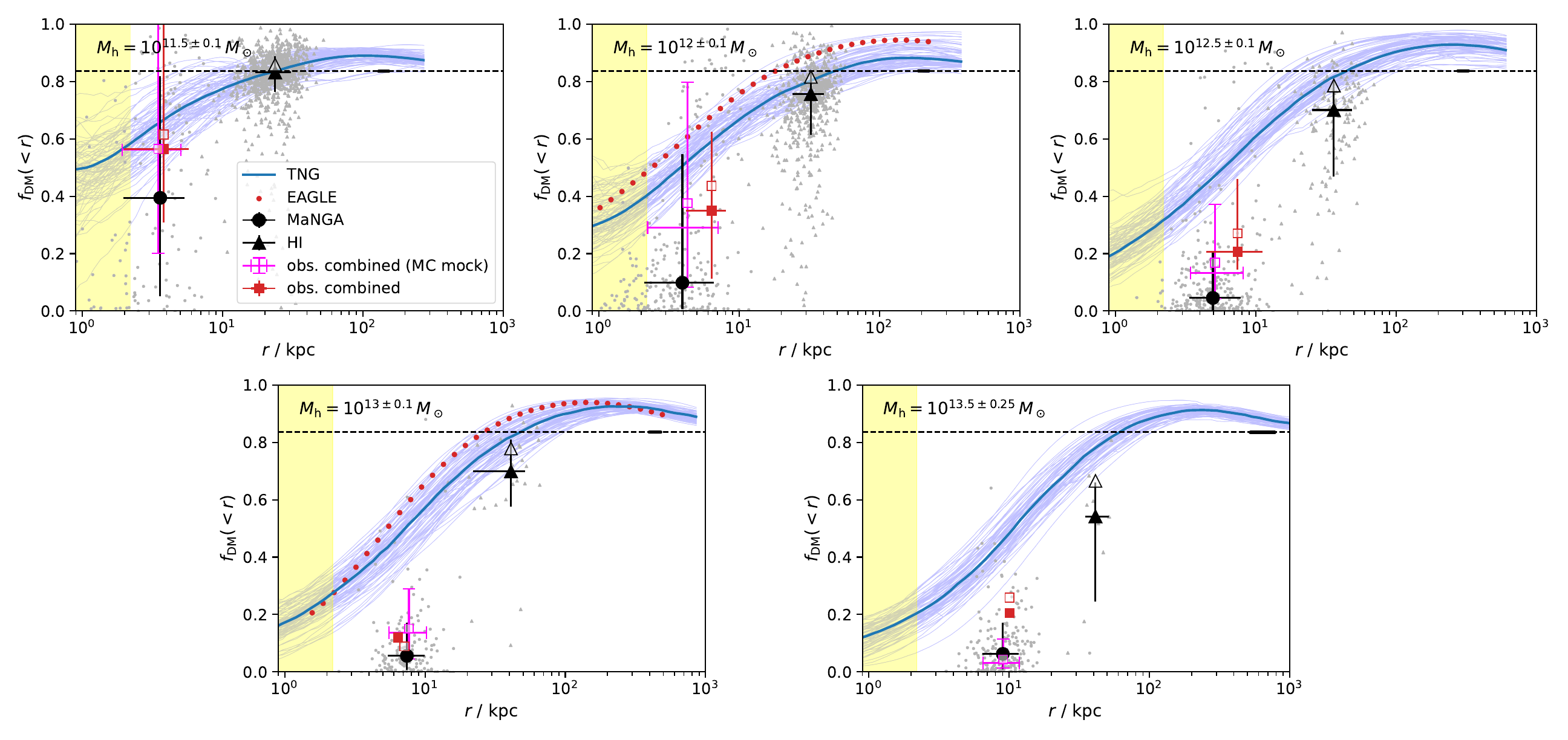}
\caption{\textbf{The impact of IMF variation on} \HI\textbf{-based dark matter fractions}.
Same as panels a--e in Fig.~\ref{fig:popproffdm}, but the stellar mass, $M_\star$, is divided by $\alpha=0.66$, accounting for the variation of different IMFs, when estimating the enclosed dark matter mass, $M_\mathrm{DM}(<r_\mathrm{HI})$. As a result, the $f_\mathrm{DM}(<r_\mathrm{HI})$ (black triangles and errorbars) and the alternative estimation of $f_\mathrm{DM}$ at radius $r_\mathrm{1/2}$ combining various datasets (magenta and red squares and errorbars; see \sref{sec:method} for details) become smaller. For comparison, the counterparts of these data in Fig.~\ref{fig:popproffdm} are indicated by hollow markers (black triangles, magenta squares, and red squares). 
The main conclusions are not significantly affected by changing the IMF.
}
\label{fig:popproffdmimfhi}
\end{figure*}

\endgroup
 \end{appendices}

\end{document}